\documentclass[12pt,aps,roupedaddress,floatfix]{revtex4}

\usepackage{color}
\usepackage{latexsym}

\usepackage{amsmath}
\usepackage{graphicx}

\usepackage{dcolumn}

\input epsf

\def\lesssim{\mathrel{\hbox{\rlap{\hbox
{\lower4pt\hbox{$\sim$}}}\hbox{$<$}}}}
 
\def\gtrsim{\mathrel{\hbox{\rlap{\hbox
{\lower4pt\hbox{$\sim$}}}\hbox{$>$}}}}

\begin{document}

\title{Simulation of astrophysical jet using the special relativistic hydrodynamics code }
	
\author{Orhan Donmez, Refik Kayali}

\address{Nigde University Faculty of Art and Science, 
Physics Department, Nigde, Turkey} 


\date{\today}

\begin{abstract}

This paper describes a multidimensional hydrodynamic code which can be 
used for the studies 
of relativistic astrophysical  flows. The code solves the special relativistic 
hydrodynamic equations as a hyperbolic system of conservation laws based on High Resolution 
Shock Capturing (HRSC) Scheme. Two standard tests, one of which is the relativistic blast wave tested
in our previous paper\cite{DO1}, and the other is the collision of two ultrarelativistic blast waves 
tested in here, are presented to demonstrate that the code captures correctly and gives
solution in the discontinuities, accurately.

The relativistic astrophysical jet is modeled for the ultrarelativistic flow case. The dynamics of jet
flowing is then determined by the ambient parameters such as densities, and velocities of the jets and 
the momentum impulse  applied to the computational surface.  We obtain solutions for the jet structure, 
propagation of jet during the time evolution, and variation in the Mach number on the computational 
domain at a fixed time.
\end{abstract}

\maketitle

\section{INTRODUCTION}

Many high-energy astrophysical problems(SNRs, $\gamma$-ray bursts -GRBs, 
Active  Galactic Nuclei (AGN) hot spots, etc.) 
involve relativistic  flows, and thus understanding 
relativistic  flows is important for interpreting the astrophysical phenomena correctly.
For instance, intrinsic beam velocities typically larger than $0.9c$ are required to explain 
the apparent superluminal motions observed in relativistic jets in microquasars in 
Galaxies\cite{MirRod} as well as in extragalactic radio sources associated with AGN.
The shocks created during these phenomena accelerate particles which emit the observed radiation. 
In particular, 
it is widely accepted that the recently discovered GRBs afterglow results from an 
emission by relativistic shocks, created by the interaction between an initial 
ejecta and the interstellar medium. The recent observations of GRB  afterglow have 
lead to numerous attempts to model these phenomena.

General relativistic effects must be considered when strong gravitational fields are 
encountered as, for example, in the case of coalescing neutron stars or near black holes. 
The significant gravitational wave signal produced by some of these phenomena can also 
only be understood in the framework of the theory of general relativity. 
Another field of research, where special relativistic flows are encountered, 
is heavy-ion collision experiments performed with large particle accelerators. 
The heavy-ions are accelerated up to ultra-relativistic velocities to study various 
aspects of heavy-ion collision physics (e.g., multi-particle production, the 
occurrence of nuclear shock waves, collective flow phenomena, or dissipative processes) 
to explore the equation of state for hot dense nuclear matter, and to find evidence for 
the existence of the quark-gluon plasma.

Multi-wavelength observation of extragalactic jets performed in radio, optical and X-ray bands 
have extended dramatically our knowledge about the complex phenomenology of these objects.
By means of these observations, different kinds of objects that can house these jets have been
characterized, at the moment we can classify distinctly these sources as radio galaxies
and quasars. A deeper analysis of these and other related objects led us to the idea of AGN.
The images obtained from Hubble Space Telescope revealed a new detail in gas flow and shock wave patterns
involving astrophysical jets and colliding interstellar winds of particles. 
Astrophysical jets are defined as highly collimated outflows in the form of high velocity 
mass flows. The outflows are observed in young stellar objects, proto-planetary nebula, 
compact objects, AGN, and GRBs\cite{Frank1} \cite{Norman1}. The jets include high Mach number
and interact with surrounding ambient gas\cite{Stone1}.

Simulating 
the fluid flow and shock wave patterns and detailed temperature profiles by implementing 
theoretical models in a gas dynamics simulator will help in analyzing the processes at work in 
these astrophysical objects. In these investigations, we apply the HRSC scheme  with
appropriate initial and boundary conditions to simulate
relativistic astrophysical jets with high Mach number from the compact objects. Pioneered in this field 
Ref.\cite{NSWS} is the pioneer in this field. They were able to show that a flow of supersonic plasma remains 
stable and develops features that could be identified with the features in the 
observations of radio galaxies. A circular and canonical deceleration area called Mach disk as
the hot spot, which is a strongly collimated beam as the elongated structure of the jets, and a big zone 
of exhaust material as the lobes of radio galaxies\cite{Ferrari}. Meanwhile, more physics 
has been included in the calculations  such as the parameters varying in the perpendicular to 
the jet flow direction, the jet radius magnetic field \cite{Clarke}\cite{Lind} and the 
special relativistic effects\cite{Komissarov}\cite{AloyIbanez}. Jets inside massive star, observed 
and called highly relativistic jets, have been studied numerically in both Newtonian\cite{Khokhlov1}
\cite{MaFadyen1} and relativistic\cite{Aloy1} \cite{Zhang1} simulations
and it has been shown that the collapsed model is able to explain many of the observed 
characteristic  of GRBs. The collapsar is formed when the iron core of a rotating massive star 
collapses to black hole and an accretion disk. The explosive deaths of massive stars produce
relativistic jets.  These studies are also declared that the 
interaction of jet with the matter at the stellar surface and the stellar wind could carry the information
for activities. The cocoon of the jet would also have different properties and the shocks within
cocoon and jet could lead to $\gamma-$ray and $x-$ray transients. 

In this paper, our goal is to simulate the ultrarelativistic jet problem using the already existing 
code which solves fully special relativistic hydrodynamics equations with HRSC scheme.

\section{FORMULATION}
\label{formulation}
A general technical description of the Hydrodynamic
equation is given by Donmez\cite{DO1} as they will be used in our code development and 
the analytic description of different problems.

\noindent
The GRH equations are written in the standard
covariant form, consist  of the local conservation laws of the
stress-energy tensor $T^{\mu \nu }$  and the matter current density $
J^\mu$:

\begin{eqnarray}
\bigtriangledown_\mu T^{\mu \nu} = 0 ,\;\;\;\;\;
\bigtriangledown_\mu J^\mu = 0.
\label{covariant derivative}
\end{eqnarray}

\noindent
Greek indices run from 0 to 3, Latin indices
from 1 to 3, and units in which the speed of light $c=1$ are
used. $\bigtriangledown_\mu$
stands for the covariant derivative with respect to the 4-metric of the
underlying spacetime, $g_{\mu \nu}$.

Defining the characteristic waves of the general
relativistic hydrodynamical equations is not trivial with imperfect
fluid stress-energy tensor. The viscosity and heat
conduction effects are neglected.  This defines the  perfect fluid
stress-energy tensor. This stress-energy tensor is used  to derive the
hydrodynamical equations. Using 
perfect fluid stress-energy tensor, we can solve some problems which
are  solved by the Newtonian hydrodynamics with viscosity, such as
those involving angular momentum transport and shock waves on an
accretion disk, etc. Entropy for 
perfect fluid is conserved along the fluid lines. The stress energy tensor
for a perfect fluid is given as

\begin{equation}
T^{\mu \nu} = \rho h u^\mu u^\nu + P g^{\mu \nu}.
\label{des 7}
\end{equation}

\noindent
A perfect fluid  is a fluid that moves through spacetime with a
4-velocity $u^{\mu}$ which may vary from event to event. It exhibits a
density of mass $\rho$ and isotropic pressure $P$ in the rest frame of
each fluid element. $h$ is the specific
enthalpy, defined as

\begin{equation} 
h = 1 + \epsilon +\frac{P}{\rho}.
\label{hdot}
\end{equation}

\noindent
Here  $\epsilon$ is the specific internal energy. The equation of
state might have  the 
functional form $P = P(\rho, \epsilon)$. The perfect
gas equation of state, 

\begin{equation}
P = (\Gamma -1 ) \rho \epsilon,
\label{flux split21}
\end{equation}

\noindent
is such a functional form.

The conservation laws in the form given in Eq.(\ref{covariant
derivative}) are not suitable for 
the use in advanced numerical schemes in $2D$. In order to carry out numerical
hydrodynamic evolutions and to
use  HRSC methods, the hydrodynamic equations after the 2+1 split
must be written as a hyperbolic system of first order flux
conservative equations. The Eq.(\ref{covariant derivative}) is written in
terms of coordinate derivatives, using the coordinates ($x^0 = t, x^i$) where and rest of the 
paper $i=1,2$ and $j=1,2$. 
Eq.(\ref{covariant derivative}) is projected onto the
basis $\lbrace n^\mu, (\frac{\partial}{\partial x^i})^\mu \rbrace$,
where $n^\mu$ is a unit timelike vector normal to a given
hypersurface. After a straightforward calculation conservation form of Special Relativistic 
Hydrodynamical (SRH) equation can be written,

\begin{equation}
\partial_t \vec{U} + \partial_x \vec{F}^x +  \partial_y \vec{F}^y = 0, 
\label{desired equation}
\end{equation} 

\noindent
where $\partial_t = \partial / \partial t$ and $\partial_x = \partial
/ \partial x^x$. This basic step serves to identify the
set of unknowns, the vector of conserved quantities $\vec{U}$, and
their corresponding fluxes $\vec{F}(\vec{U})$. With the
equations in conservation form, almost every high
resolution method devised to solve hyperbolic systems of conservation
laws can be extended to SRH.

The evolved state vector $\vec{U}$ consists of  the conservative
variables $(D, S_x, S_y, \tau)$ which are conserved variables for density,
momentum in $x$ and $y$ direction and energy respectively; in terms of the
primitive variables $(\rho, v^x, v^y , \epsilon)$, this becomes,

\begin{equation}
D = \sqrt{\gamma} W \rho ;\;\;\
S_x = \sqrt{\gamma}\rho h W^2 v_x ;\;\;\
S_y = \sqrt{\gamma}\rho h W^2 v_y ;\;\;\
\tau  = \sqrt{\gamma} (\rho h W^2 - P - W \rho)
\label{matrix form of conserved quantities}
\end{equation}

\noindent
Here $\gamma$ is the determinant of the 3-metric $\gamma_{ij}$, which is unitary matrix, 
$v_x$ and $v_y$ are the fluid 3-velocity, the lapse function$\alpha=1$  and the shift vector $\beta=0$ 
for SRH and W is the Lorentz factor,

\begin{equation}
W = \alpha u^0 = (1 - v^i v^j)^{-1/2}. 
\label{Wdot1}
\end{equation}

\noindent 
The flux vectors $\vec{F^i}$ are given by

\begin{equation}
\vec{F}^i =  \left( \begin{array}{c}
v^i D \\
v^i S_j + P \delta ^{i}_{j}   \\
(v^i \tau + v^i P  \end{array} \right). 
\label{matrix form of Flux vector}
\end{equation}

\noindent
The spatial components of the 4-velocity $u^i$ are related to the
3-velocity by the following formula: $u^i = W v^i$. The source vector $\vec{S} =0$ for SRH case.


\subsection{Spectral Decomposition and Characteristic Fields}
\label{Spectral Decomposition and Characteristic Fields}

The use of HRSC  schemes requires the
spectral decomposition of the Jacobian matrix of the
system, $\partial \vec{F^i} / \partial \vec{U}$. 
The spectral decomposition of the Jacobian matrices of the SRH 
 equations with a general equation of state
was reported here . It is displayed the 
spectral decomposition, valid for a generic spatial metric, 
in the x-direction, $(\partial \vec{F}^x / \partial \vec{U})$;
permutation of the indices yields the $y$ direction.   

We started the solution by considering an equation of state in which the
pressure $P$ is a function of $\rho$ and $\epsilon$, $P = P(\rho,
\epsilon)$. The relativistic speed of sound in the fluid $C_s$ is
given by

\begin{equation}
C_{s}^2 = \left. \frac{\partial P}{\partial E} \right|_S. =  \frac{\chi}{h} + \frac{P
\kappa}{\rho^2 h},   
\label{cdot}
\end{equation}

\noindent
where $\chi = \partial P / \partial \rho\vert_\epsilon$, $\kappa = \partial
P / \partial \epsilon\vert_\rho$, $S$ is the entropy per particle, and
$E = \rho + \rho \epsilon$ is the total rest energy density.


A complete set of the right eigenvectors $[\vec{r}_i]$ and  corresponding
eigenvalues $\lambda_i$  along the $x$-direction obeys

\begin{eqnarray}
\left[ \frac{\partial \vec{F}^x}{\partial \vec{U}} \right] \left[
\vec{r_i} \right] = 
\lambda_i \left[ \vec{r_i} \right], \; \; \; i= 1,...,4.
\label{eigenvalue-vector equation}
\end{eqnarray}

\noindent
The solution contains triply degenerate eigenvalues,

\begin{eqnarray}
\lambda_1^x = \lambda_2^x = \lambda_3^x = v^x .
\label{eigenvalue123}
\end{eqnarray}

\noindent
The other eigenvalues are given as

\begin{eqnarray}
\lambda_{\pm}^x = \frac{1}{1-v^2 C_s^2} \Big\{v^x (1-C_s^2) \pm
\nonumber \\
\sqrt{C_s^2 (1-v^2) [(1-v^2 C_s^2) - v^x v^x (1 - C_s^2)]}
\Big\}.
\label{eigenvalue plus minus}
\end{eqnarray}

\noindent
A set of linearly independent right eigenvectors spanning  this space
is given as 

\begin{eqnarray}
\vec{r}_1^x = \left[ \frac{\kappa}{h W (\kappa - \rho C_s^{2})}, v_x, v_y,
1 - \frac{\kappa}{h W (\kappa - \rho C_s^{2})} \right]^T,
\label{eigenvector1}
\end{eqnarray}

\begin{eqnarray}
\vec{r}_2^x = \bigg[ W v_y, h 2 W^2 v_x v_y, h (1 +
2 W^2 v_y v_y), v_y W (2 W h -1)\bigg]^T,
\label{eigenvector2}
\end{eqnarray}

\noindent
and

\begin{eqnarray}
\vec{r}_{\pm}^x = \bigg[1, h W (v_x - \frac{v^x -\lambda_{\pm} 
}{1 - v^x \lambda_{\pm}}, h W
v_y, \frac{h W (1 -v^x v^x)}{1 -v^x \lambda_{\pm}} - 1 \bigg]^T. 
\label{eigenvector plus minus}
\end{eqnarray}

\noindent
Here, the superscript T denotes the transpose.

Now, the eigenvalues and eigenvectors are computed in the $y$-direction using the 
information in the $x$-direction in Donmez \cite{DO1} for general relativistic 
hydrodynamic equations  .
Let ${\bf M}^1(U)$ be the matrix of the right eigenvectors, that is, the
matrix having as columns the right eigenvectors with the standard
ordering(\cite{DO1}, \cite{BFIMM}), 
${\bf M}^1 = (\vec{r}_-, \vec{r}_1, \vec{r}_2,
\vec{r}_+)$.  To obtain the spectral decomposition in the
other spatial directions $q = 2 (\equiv y)$, it is enough to take
into account the following symmetry relations:

\newcounter{abeann}
\begin{list}%
{\arabic{abeann})}{\usecounter{abeann}
	\setlength{\rightmargin}{\leftmargin}}

\item The eigenvalues are easily obtained by carrying out in equations
(\ref{eigenvalue123}) and (\ref{eigenvalue plus minus}) a simple
substitution of indices $x$ by $q$. 

\item The matrix $M^1$ in the $x$-direction is permuted according to 
${\bf M^q} = {\cal P}_{q1}({\bf M}^1)$,
where the two operators ${\cal P}_{q1}$ act on their arguments by the
following sequential operations:

\begin{itemize}

\item to permute the second and $(q + 1)$th rows,

\item to interchange indices $1 \equiv x$ with $q$.

\end{itemize}

\end{list}

From Donat et. al.\cite{DFIM}, the left
eigenvectors in the $x$-direction are :

\begin{eqnarray}                                                      
{\bf l}_{0,1} = {\displaystyle{\frac{W}{{\cal K} - 1}}} 
\left[ \begin{array}{c}
h - W
\\  \\
W v^x
\\  \\
W v^y
\\  \\
- W
\end{array} \right]
\end{eqnarray}

\begin{eqnarray}                                                     
{\bf l}_{0,2} = {\displaystyle{\frac{1}{h \xi^x}}}
\left[ \begin{array}{c}
- v_y 
\\  \\
v^x v_y 
\\  \\
(1 - v_x v^x)  
\\  \\
-  v_y 
\end{array} \right]
\end{eqnarray}

\begin{eqnarray} 
{\bf l}_{\mp} = ({\pm} 1){\displaystyle{\frac{h^2}{\Delta^x}}} 
\left[ \begin{array}{c} 
h W {\cal V}^x_{\pm} \xi^x + (\pm \frac{\Delta}{h^2}) l_{\mp}^{(5)} 
\\  \\
\Gamma_{xx} (1 - {\cal K} {\tilde {\cal A}}^x_{\pm}) +
(2 {\cal K} - 1){\cal V}^x_{\pm} (W^2 v^x \xi^x - \Gamma_{xx} v^x)
\\  \\
\Gamma_{xy} (1 - {\cal K} {\tilde {\cal A}}^x_{\pm}) +
(2 {\cal K} - 1) {\cal V}^x_{\pm} (W^2 v^y \xi^x - \Gamma_{xy} v^x)
\\  \\
(1 - {\cal K})[- \gamma v^x + {\cal V}^x_{\pm} (W^2 \xi^x - \Gamma_{xx})]
- {\cal K} W^2 {\cal V}^x_{\pm} \xi^x 
\end{array} \right]
\end{eqnarray}   


\noindent
The variables used for the left eigenvectors in the $x$-direction are defined as follows:

\begin{eqnarray*}
{\cal K} \equiv {\displaystyle{\frac{\tilde{\kappa}}
{\tilde{\kappa}-c_s^2}}} \,\,\,\,\,,\,\,\,\,\,
\tilde{\kappa}\equiv \kappa/\rho 
\end{eqnarray*}

\begin{eqnarray*}
{\cal C}^x_{\pm} \equiv v_x - {\cal V}^x_{\pm} \,\,\,,\,\,\,
{\cal V}^x_{\pm} \equiv {\displaystyle{\frac{v^x - \lambda_{\pm}^x}
{\gamma^{xx} - v^x \lambda_{\pm}^x}}} 
\end{eqnarray*}

\begin{eqnarray*}
{\tilde {\cal A}}^x_{\pm} \equiv 
{\displaystyle{\frac{1 - v^x v^x}
{1 - v^x {\lambda}^x_{\pm}}}}
\end{eqnarray*}

\begin{eqnarray*}
1 - {\tilde {\cal A}}^x_{\pm} = v^x {\cal V}^x_{\pm} \,\,\,,\,\,\,
{\tilde {\cal A}}^x_{\pm} - {\tilde {\cal A}}^x_{\mp} =
v^x ({\cal C}^x_{\pm} - {\cal C}^x_{\mp}) 
\end{eqnarray*}

\begin{eqnarray*}
({\cal C}^x_{\pm} - {\cal C}^x_{\pm}) +
({\tilde {\cal A}}^x_{\mp} {\cal V}^x_{\pm} -
{\tilde {\cal A}}^x_{\pm} {\cal V}^x_{\mp}) = 0 
\end{eqnarray*}

\begin{eqnarray*}
\Delta^x \equiv h^3 W ({\cal K} - 1) ({\cal C}^x_{+} - {\cal C}^x_{-}
) \xi^x 
\end{eqnarray*}

\begin{eqnarray*}
\xi^x \equiv 1  -  v^x v^x 
\end{eqnarray*}

Detailed informations related with numerical solution of GRH equations using high
resolution shock capturing schemes in $3D$ are given in our paper Donmez\cite{DO1}.
This reference explains detail description of numerical solutions of SRH equations,
Adaptive-Mesh Refinement, high resolution shock capturing method used,
and solution of GRH equation using  Schwarzschild coordinate as a source term. The code is
constructed for general spacetime metric with lapse function and shift vector and is fully
parallelized to make optimum use of supercomputers which is necessary to achieve the
numerical modeling of real astrophysical problems, such as astrophysical jet problems, 
coalescing of the compact binaries and accretion disk around the compact objects.


\section{Numerical Results}
\label{Numerical Results}

\subsection{Boundary Condition}
\label{Boundary Condition}

Boundary conditions are set by filling the data in guard cells  with
appropriate values. In the numerical calculation, boundary filling
plays an important role in the simulations. The computational grid is
extended at both sides of the physical domain to compute the fluxes at
interfaces. These extra cells are also called guard cells or ghost
zones. In this paper, we have used outflow boundary condition at each ghost zone.
This boundary condition has to be provided on each time step for all primitive and
conservative variables in the special relativistic hydro code.
The outflow boundary condition for the computational domain is as follows. For velocity:
$U_{0}^{n} = -\frac{1}{2}(|U_{1}^{n}| - U_{1}^{n})$ and 
$U_{M+1}^{n} = \frac{1}{2}(|U_{M}^{n}| + U_{M}^{n})$. For the other variables:
$U_{0}^{n} = U_{1}^{n}$ and $U_{M+1}^{n} = U_{M}^{n}$, where $n$ represents the time step.

\subsection{Relativistic Blast Wave}
\label{Relativistic Blast Wave}

Riemann problems with large initial pressure jumps produce blast waves with dense 
shells of material propagating at relativistic speeds. The Riemann shock tube
is a useful test problem because it has an exact time-dependent
solution, and tests the ability of the code to evolve both smooth and
discontinuous flows. 
For appropriate initial conditions, both the speed of the leading shock front and 
the velocity of the shell material approach the speed of light producing very narrow 
structures. The accurate description of these thin, relativistic shells involving 
large density contrasts is a challenge for any numerical code. Some particular blast 
wave problems have become standard numerical tests. 
One of the common test problem is called the special relativistic shock tube problem. 
Testing the code with this problem is given our paper Donmez\cite{DO1}.

\subsection{Collision of two relativistic blast waves in 2D}
\label{Collision of two relativistic blast waves in 2D}

The collision of two strong blast waves in one-dimension was used by 
Woodward et al.\cite{WCol} to compare the performance of 
several numerical methods in classical hydrodynamics. In the relativistic case, 
Yang et al.\cite{YCTC} considered the same problem to test the high-order extensions 
of the relativistic beam scheme, whereas Marti et al.\cite{MaMu} used it to evaluate the 
performance of their relativistic PPM code. In the last treatment, the original boundary 
conditions were changed (from reflecting to outflow) to avoid the reflection and 
subsequent interaction of rarefaction waves allowing for a comparison with an analytical solution. 
The initial data corresponding to this test, consisting of three constant states with 
large pressure jumps at the discontinuities separates the states.

\begin{center}
\begin{figure}
\vspace*{2.2cm}
\centerline{\epsfxsize=14cm \epsfysize=14cm
\epsffile{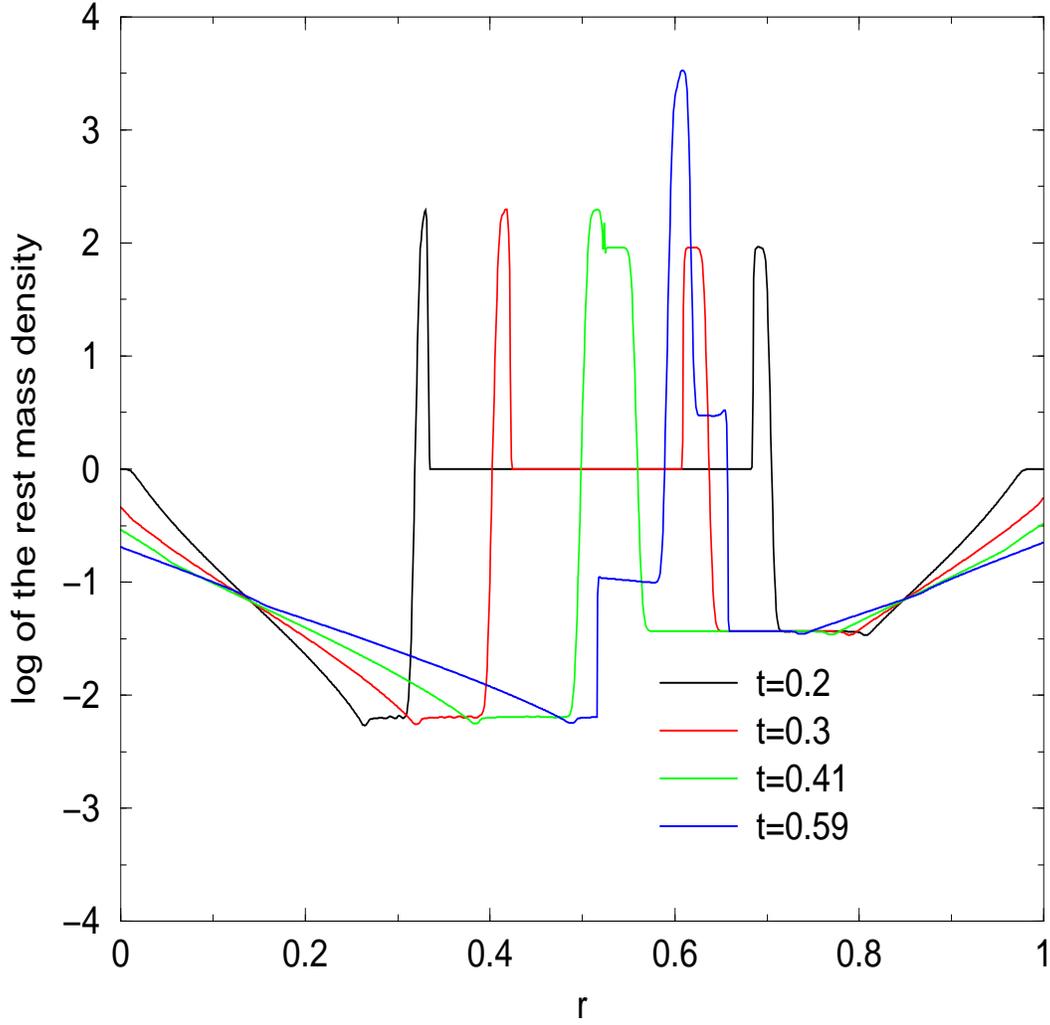}}
\caption{The evolution of log of the density distribution for the colliding relativistic 
blast wave problem up to the interaction of the waves. The data is taken from the 
two-dimensional colliding wave extracting from the diagonal axis.
The computation has been performed with relativistic hydrodynamical code  on an equidistant 
grid of 1024 zones}
\label{density diff time}
\end{figure}
\end{center}

\begin{center}
\begin{figure}
\vspace*{2.2cm}
\centerline{\epsfxsize=14cm \epsfysize=14cm
\epsffile{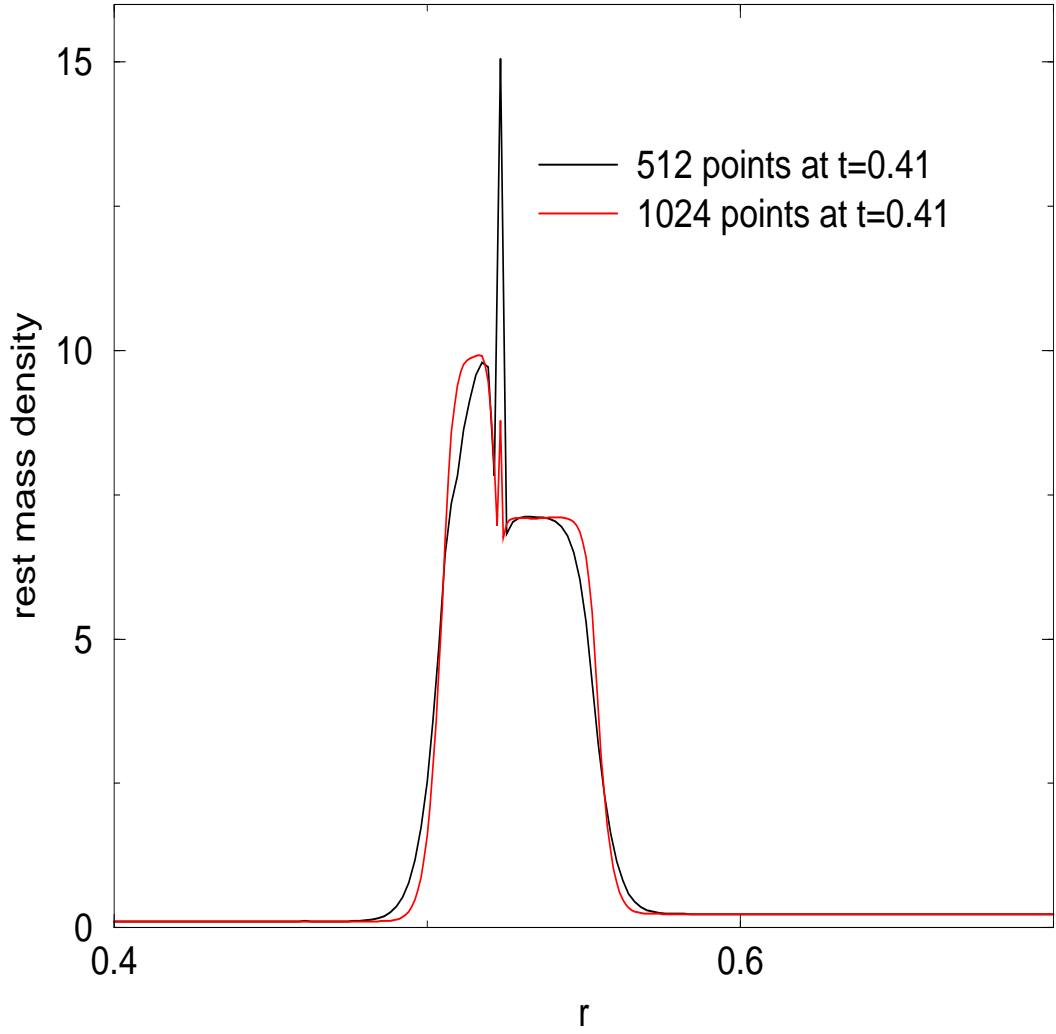}}
\caption{The evolution of density at $t=0.41$ at different resolution. The preshock and postshock 
region are resolved with more grid zones.}
\label{density at different resolution}
\end{figure}
\end{center}

\begin{center}
\begin{figure}
\vspace*{2.2cm}
\centerline{\epsfxsize=14cm \epsfysize=14cm
\epsffile{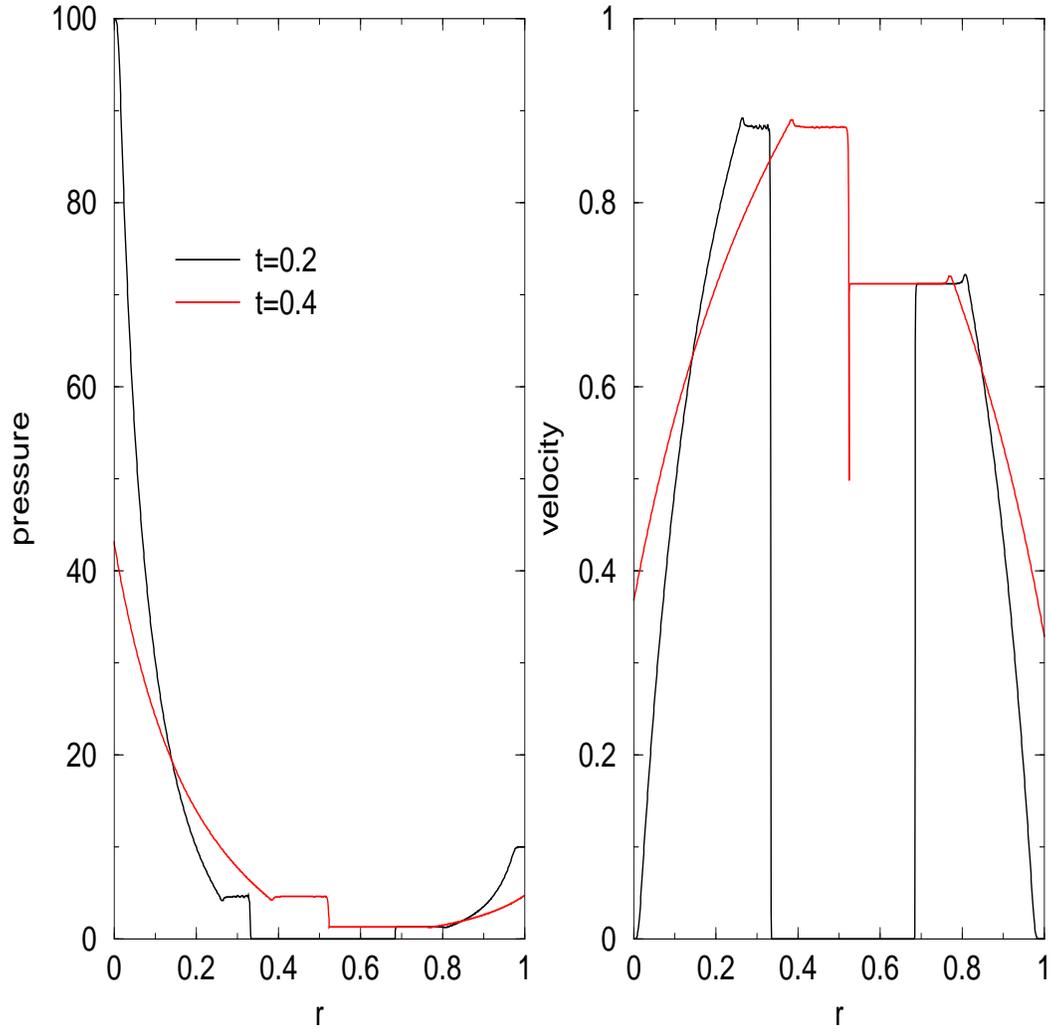}}
\caption{The left pane: the evolution of pressure at two-different time. Right panel: the 
evolution of velocity of the blast wave.}
\label{Pressurevelocity diff time}
\end{figure}
\end{center}

In this report, the initial data corresponds to collision of two strong blast waves in $2D$, consisting 
of three constant states with large pressure jumps at the discontinuities separating the 
states (at $r = 0.15$ and $r = 0.85$, where $r=sqrt(y^2 + z^2)$), as well as containing the properties 
of the blast waves created by the decay of the initial discontinuities which are located
on the main diagonal of the square box and the data is extracted along the other diagonal axis.
The initial data for this test is listed in 
Table\ref{Initial data for two relativistic blast wave collision}.  The initial pressure 
discontinuities drive shocks into the middle part of the grid; behind them, rarefaction form and
propagate toward the outher boundaries where they are falling out from computational domain.
When the two shock waves collied at $t=0.41$, a very dense shell is created. 
The collision gives rise to a narrow region of very high density as seen in 
Fig.\ref{density diff time} bounded by two shocks moving at speeds $0.88$ (shock at the left) 
and $0.711$ (shock at the right). The velocity of shock wave during the time evolution is plotted 
in the right panel of Fig.\ref{Pressurevelocity diff time}.  
The presence of very narrow structures involving large 
density jumps requires very fine zoning and high resolution scheme to resolve the states properly.
The high resolution shock capturing scheme we used here  resolves the structure of the 
collision region satisfactorily well.
Fig.\ref{density at different resolution} shows that more resolutions 
are needed to resolve the preshock and postshock states reasonably well. The left panel at 
Fig.\ref{Pressurevelocity diff time} shows the behavior of pressure jump during the time evolution.
The outflowing shock does not propagate along the grid and the contact discontinuity is not created on the 
numerical results.

\subsection{Astrophysical Jet}
\label{Astrophysical Jet}

Investigation of gas flow in close binary systems is important in considering the accretion disk
around the compact objects. The interaction of the disk with the compact objects, through 
processes that are still poorly understood, produces jets with high energy to mass ratio. Here 
we are primarily concerned  with the propagation of these jets, not so much with how they are born. 
The propagating jets produce solar wind in the medium  and this wind is highly supersonic. 
So obstacles in the flow, large variations in stream 
speeds, or fast ejecta produce shock waves, just as shock waves are produced by a 
supersonic jet. With a jet, sound waves develop ahead of the jet nose because the 
matter is traveling faster than the waves can escape. A shock wave forms, 
due to pressure buildup, to keep the air flowing around the jet. 

Jets are, by current theory, the best, perhaps, essential method of explaining how such large amounts 
of energy can be observed from objects at cosmological distances. However, there is much speculation 
as to the exact nature of these jets, the primary questions being: how do they arise and what is 
their structure. In this paper, we focus on the question, what is the structure of GRB jets?

\begin{center}
\begin{figure}
\vspace*{2.2cm}
\centerline{\epsfxsize=14cm \epsfysize=14cm
\epsffile{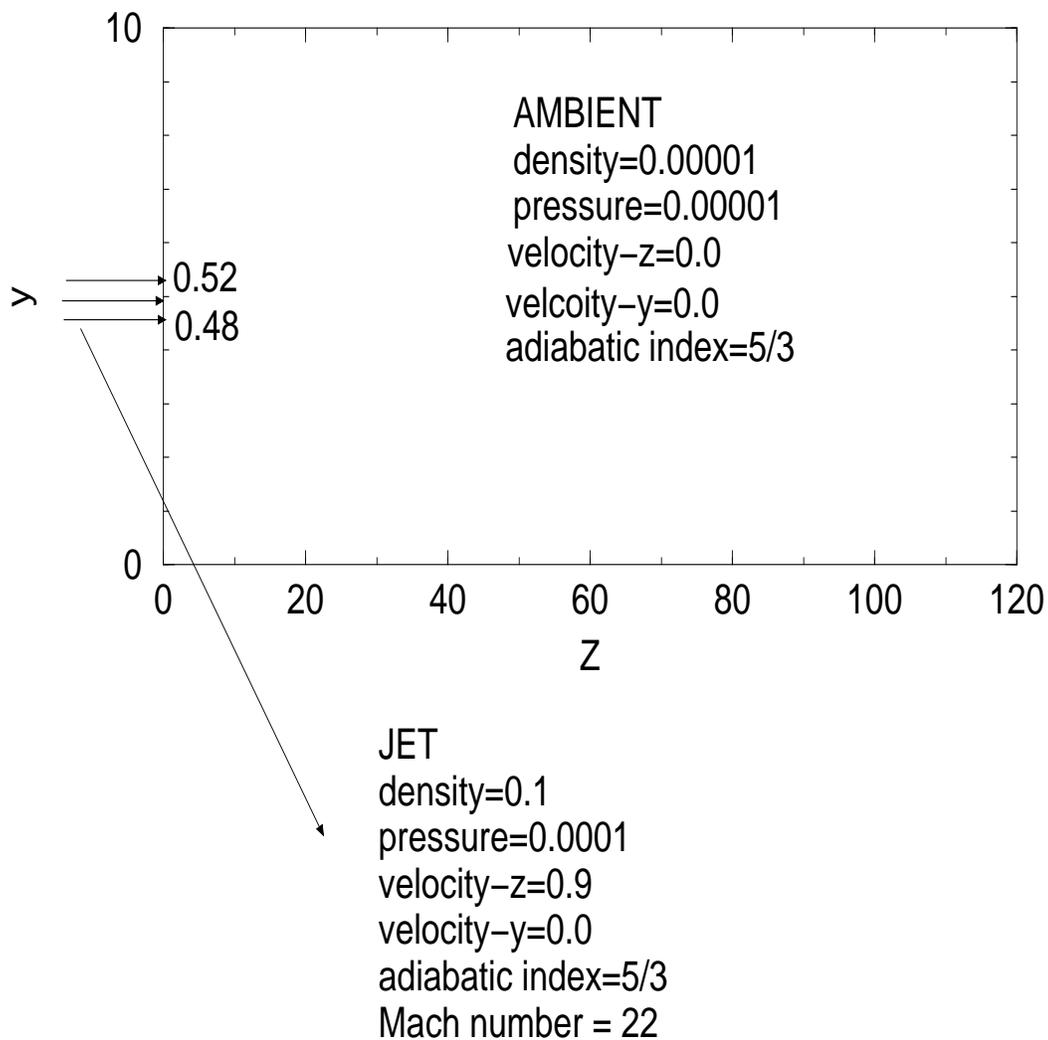}}
\caption{Initial representation of computational domain for astrophysical jet with the 
appropriate initial conditions.}
\label{Inital Setup}
\end{figure}
\end{center}

\begin{center}
\begin{figure}
\vspace*{2.2cm}
\centerline{\epsfxsize=15cm \epsfysize=15cm
\epsffile{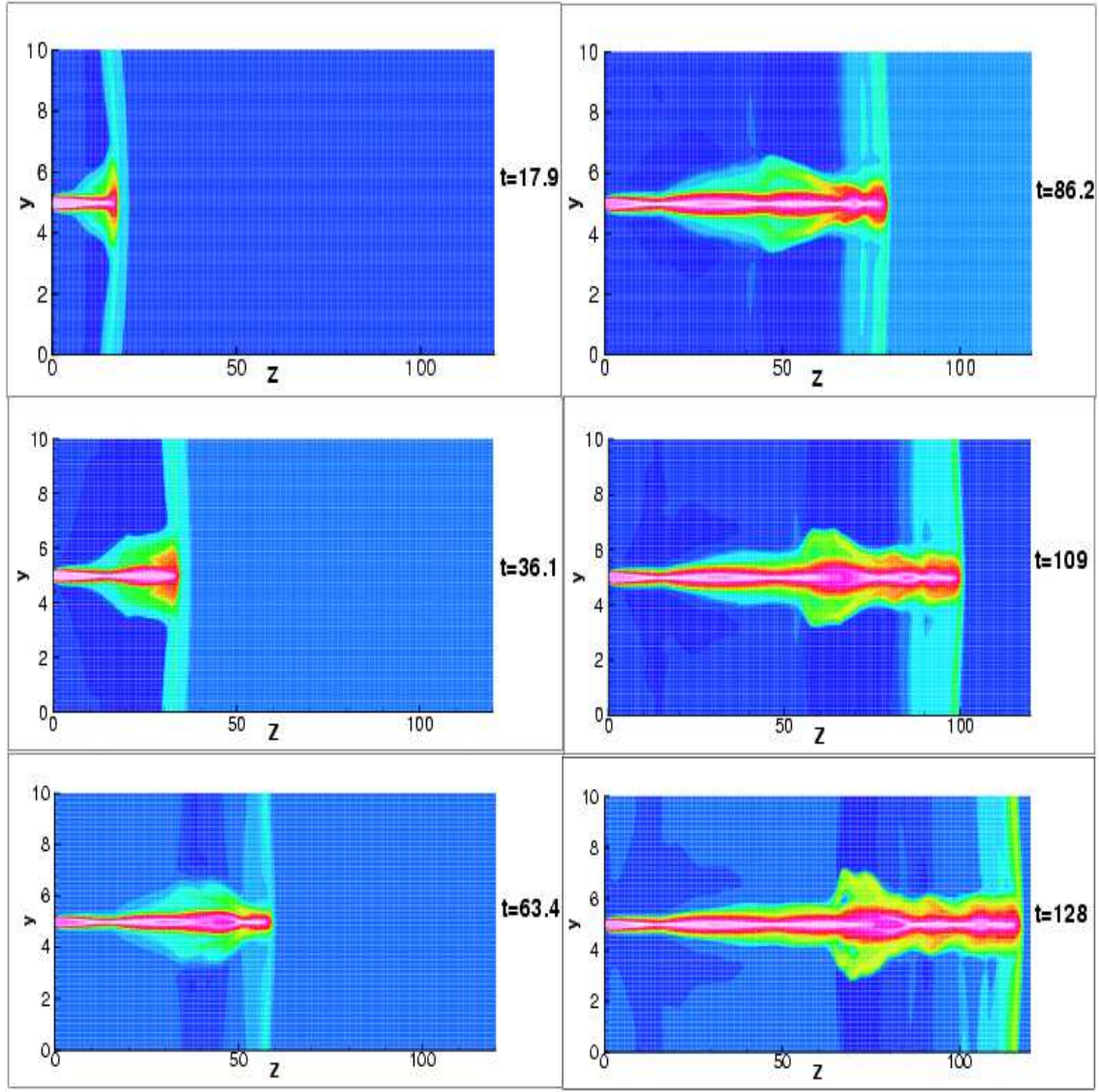}}
\caption{Contour maps of the logarithm of the rest-mass density for  relativistic jet, 
moving with $v=0.9c$, at different times. 
Snapshots of the proper rest-mass density distribution are plotted at $y-z$ plane. $256$ resolutions
in $y$ and $4096$ resolutions in $z$ are used.}
\label{2D Jet different time}
\end{figure}
\end{center}

\begin{center}
\begin{figure}
\vspace*{2.2cm}
\centerline{\epsfxsize=15cm \epsfysize=15cm
\epsffile{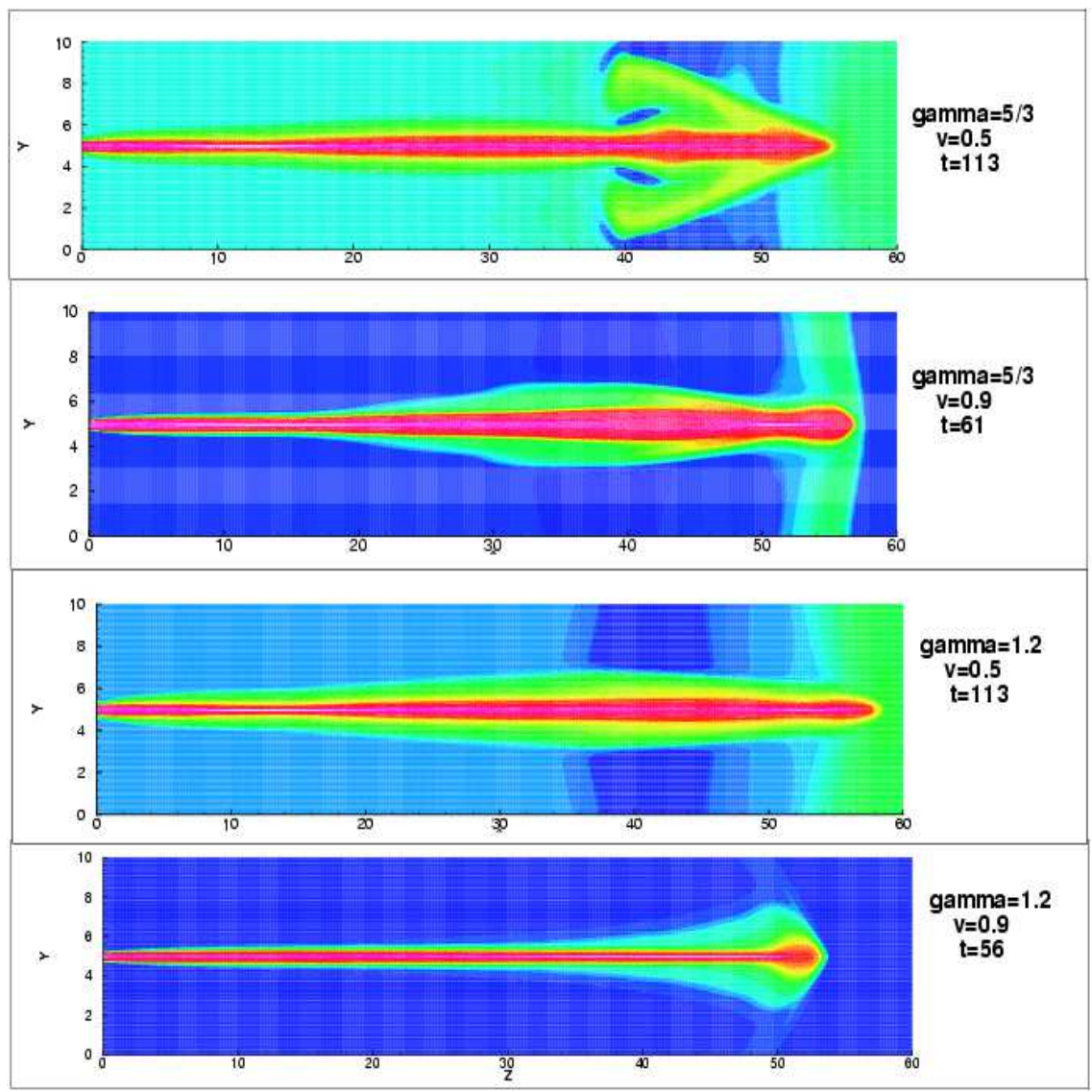}}
\caption{2D representation of jet depend of adiabatic index at a fixed time for 
different jet velocities}
\label{2D differnet gamma and velocity}
\end{figure}
\end{center}

Here, we numerically model the $2D$ simulation of a relativistic jet propagating through an 
homogeneous atmosphere. The initial computational domain and values of beam  for the flow velocity,
Mach number, rest-mass density and adiabatic index are given in Fig.\ref{Inital Setup}. The fluid 
viscosity and thermal conductivity are neglected. The ambient
medium is has a size of $10x120$. The jet is injected at $4.8<y<5.2$ in the direction of 
positive $z-$axis. We have constructed a slab jet, which is periodic in the $x-$ direction. 
An ideal gas equation of state with adiabatic index $\gamma = 5/3$ is assumed 
to describe both the jet matter and the ambient gas.  In our calculation, $256$ resolution in $y$
and $4096$ resolution in $z$  are used for 
Figs.\ref{2D Jet different time},\ref{2D differnet gamma and velocity}.  It is well known that 
the propagation of a supersonic  jet is governed by the interaction of jet matter with ambient 
medium, which  produces a bow shock in the ambient medium. The evolution of the jet was simulated
up to the time when the head of the jet is about to reach to the boundary, which is far away
from the injection point.     

The time evolution of relativistic jet for rest-mass density 
in the plane $x = 0$ is given in Fig.\ref{2D Jet different time} at different snapshots 
when the jet propagates through $y-z$ plane. The presence of emitting matter moving at 
different velocities and orientations could lead to local variations of apparent relativistic
motion within the jet. The distribution of apparent motions during the time evolutions is
inhomogeneous and fringed as seen in Fig.\ref{2D Jet different time}. The structure of jet 
contains cocoon and vortexes with a bow shock. The cocoon contains shocked jet material deflected 
backward at the head of the jet. Since the jet is narrowly collimated, its beam is very thin.

\begin{center}
\begin{figure}
\vspace*{2.2cm}
\centerline{\epsfxsize=15cm \epsfysize=15cm
\epsffile{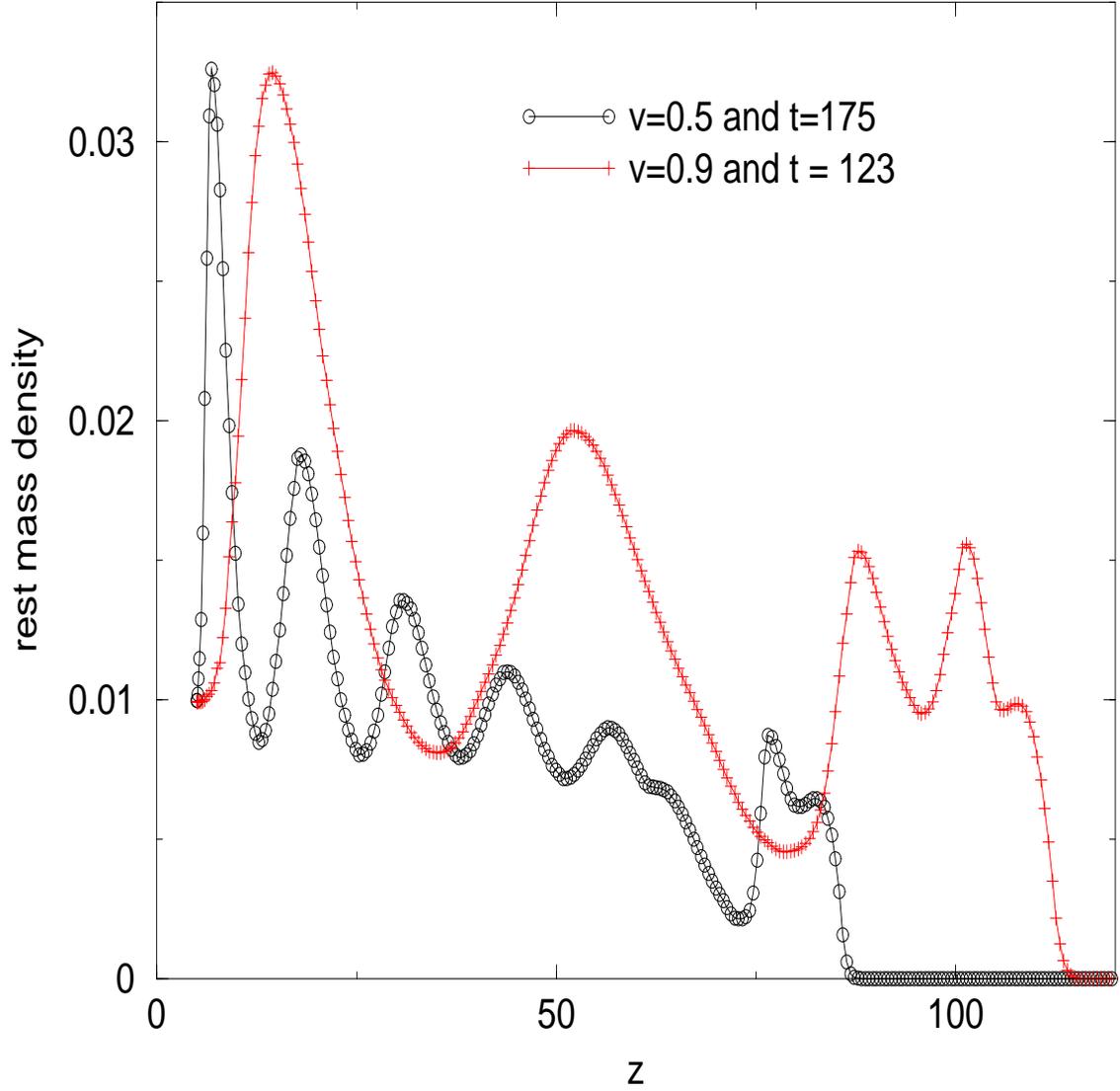}}
\caption{Vertical configuration values of rest-mass density along a cut to the z-axis, at $r=5.0$,
are shown for different jet velocity(mildly and ultrarelativistic cases). This figure illustrates
the value of rest mass density during the evolution of jet flow at fixed time. Oscillatory 
jet rest mass density is achieved.}
\label{Box120 Jet Density Diff Time}
\end{figure}
\end{center}

\begin{center}
\begin{figure}
\vspace*{2.2cm}
\centerline{\epsfxsize=15cm \epsfysize=15cm
\epsffile{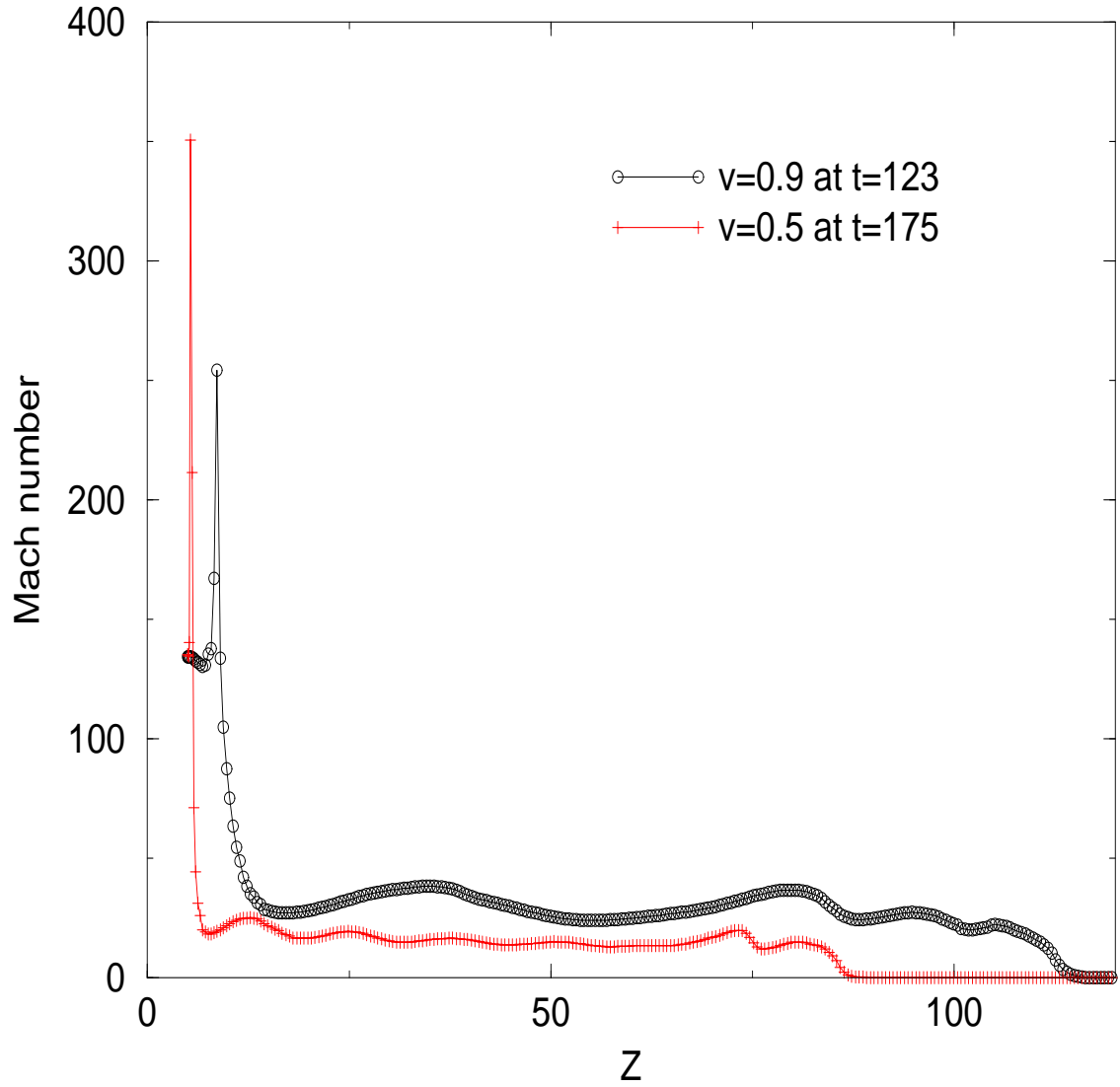}}
\caption{Behavior of Mach number as a function of Z  for different initial jet velocities 
at different times. It is taken at a fixed y=5.0}
\label{Mach number}
\end{figure}
\end{center}

\begin{center}
\begin{figure}
\vspace*{2.2cm}
\centerline{\epsfxsize=15cm \epsfysize=15cm
\epsffile{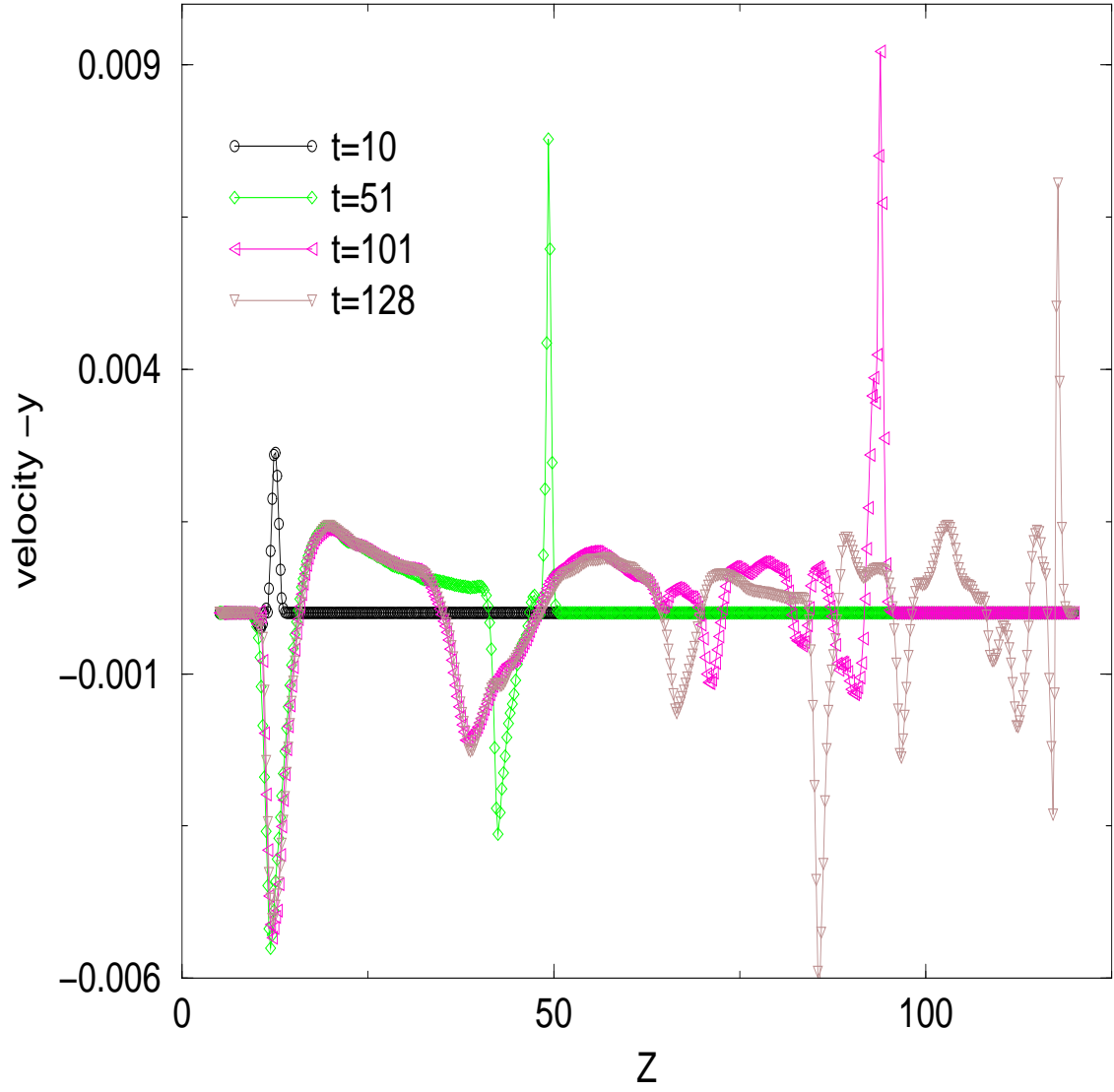}}
\caption{The $y$ velocity of relativistic jet at  $y=5$  at different times as a function of $Z$.}
\label{velocity in y}
\end{figure}
\end{center}

The Fig.\ref{Box120 Jet Density Diff Time} shows the variations in one-dimensional cut of rest 
mass density in the jet propagation direction, $z$, at final snapshot for two different jet velocities, 
$v=0.5$ and $v=0.9$. While the one with smaller velocity is called as a mildly relativistic 
jet, the other one is called the ultra-relativistic jet. The ultra-relativistic one displays rich 
internal structure with oblique shocks effectively decelerating the flow in the beam from 
Mach number equal to $140$ at the injection point down to a value of about $10$ around the 
head as seen in Fig.\ref{Mach number}.
The one-dimensional cut of the rest-mass density as given in Fig.\ref{Box120 Jet Density Diff Time}
shows that the jet density propagates 
and oscillates on the working surface and it produces time varying waves and shock waves  
which interact with ambient medium.  The pressure
is larger at the head of jet and it causes the jet to propagate through the ambient medium at a 
larger speeds in both direction. The graphs seen in Fig.\ref{velocity in y} show the variation of 
the $y$ velocity of cocoon, propagating through computational domain, at a fixed $y=5$ along 
the $z-$axis  for four different times. The head of jet has the largest $y$ velocity which 
is represented with peaks as seen in Fig.\ref{velocity in y}. 
Figs.\ref{2D Jet different time}, \ref{Box120 Jet Density Diff Time} 
and  \ref{Mach number} also depict that there is a Mach shock at the jet head, and material that has 
passed through the head is slowly forming a backflow along the jet, building up the a shear layer.

We have also examined the propagation of two-dimensional relativistic jet in case of the different initial 
velocities  and adiabatic values, $\Gamma$. The rest mass density for different $\Gamma$ values are
plotted in Fig.\ref{2D differnet gamma and velocity}. In agreement with previous studies on 
astrophysical problem, the relativistic jet is also more tightly packed with the smaller $\Gamma$. 
The lower $\Gamma$ means cooler jet with larger Mach number of the flow.  The cooler jet presents also 
a complex structure of internal shocks generated by pressure mismatches between the beam and the 
overpressured cocoon and by perturbations of the beam boundary by vortices and bulk motions within 
the cocoon. The cocoon is mainly formed by large vortices in cold jets with $\Gamma = 5/3$ while 
the strong beam collimation causes a large acceleration of the jet in cold jets with $\Gamma = 4/3$. 
The beam gas is less efficiently redirected into the cocoon, and thinner cocoon with smaller 
vortices form with $\Gamma = 4/3$ as seen in Fig.\ref{2D differnet gamma and velocity}.

\section{CONCLUSION}

We have described the main futures of stable $2D$ special relativistic hydrodynamical code and
discussed test problems  in Ref.\cite{DO1} and here involving strong shocks in two-dimension. 
The test problem solved here is called the collision of relativistic blast wave. The 
promising results from the test problem give us encouragement to simulate ultrarelativistic gas flows.

We have carried out simulations of fully relativistic astrophysical jet in $2D$. By 
detailed examination of a time series of ultrarelativistic flow, we have described the propagation 
of jet through the computational domain. The ultrarelativistic flow is thought to occur and to play a 
crucial role in the generation of GRBs. We have also computed the dynamic of jet in case of different 
adiabatic index used in equation of state. The jet with $\Gamma = 4/3$ form thinner cocoon with smaller 
vortices and the matter collimation causes a large acceleration of the jet.


\begin{acknowledgments}
This project is supported by N.U. 2003/01. It has been performed using TUBITAK/ULAKBIM super 
computers/Beowulf cluster.
\end{acknowledgments}

\newpage
\begin{table} 
\caption{Initial data for the two relativistic blast wave collision test problem in $2D$. 
The decay of the initial discontinuities (at $r = 0.15$ and $r = 0.85$) produces 
two shock waves (velocities $v_{shock}$) moving in opposite directions. The gas is 
assumed to be ideal with an adiabatic index $\gamma=5/3$. }
$$  \vbox{ \offinterlineskip \vskip7pt
  \def\qq{\hskip1.0em}
  \def\laststrut{\vrule depth6pt width0pt}
  \def\titlestrut{\vrule height10pt depth6pt width0pt}
\halign {#\vrule\strut &\quad#\hfil
       &&\qq#\vrule &\qq\hfil#\hfil \cr
\noalign{\hrule}
 & \multispan{7}\hfil Blast wave collision test problem \hfil\titlestrut 
  &\cr \noalign{\hrule}  
&  && Left && Middle && Right &\cr
 \noalign{\hrule}
& $\rho$ && 1.0 && 1.0 && 1.0 &\cr
 \noalign{\hrule}
& $p$ && 100.0 && 0.01 && 10.0 &\cr 
 \noalign{\hrule}
& $v_x$ && 0.0 && 0.0 && 0.0 &\cr 
\noalign{\hrule}
& $v_y$ && 0.0 && 0.0 && 0.0 \laststrut &\cr
\noalign{\hrule\vskip4pt}
  \noalign{\vskip5pt} }} $$ 
\label{Initial data for two relativistic blast wave collision}
\end{table}

\newpage

\end{document}